\documentclass[a4paper,12pt]{article}
\pdfoutput=1 

\usepackage{jcappub} 
\usepackage[symbol]{footmisc}
\usepackage{changes}
\usepackage{amsmath,amssymb}
\usepackage{graphicx}
\usepackage{adjustbox}
\usepackage{multirow}
\usepackage[normalem]{ulem}
\usepackage{bm}         
\newcommand{\mU}{{\mathcal{U}}}

\newcommand{\fpbh}{f_{\mathrm{PBH}}}

\begin{document}

\title{Confronting sound speed resonance with pulsar timing arrays}

\author{Jia-Heng~Jin,$^{a,b}$}
\author{Zu-Cheng~Chen,$^{c,d,e}$}
\author{Zhu~Yi,$^{d}$}
\author{Zhi-Qiang~You,$^{c,d,f}$}
\author{Lang~Liu,\note{Corresponding author}$^{c,d,*}$}
\author{You~Wu$^{g,e,*}$}

\affiliation{$^a$CAS Key Laboratory of Theoretical Physics, Institute of Theoretical Physics, Chinese Academy of Sciences, Beijing 100190, China}
\affiliation{$^b$School of Physical Sciences, University of Chinese Academy of Sciences, No. 19A Yuquan Road, Beijing 100049, China}
\affiliation{$^c$Department of Astronomy, Beijing Normal University, Beijing 100875, China}
\affiliation{$^d$Advanced Institute of Natural Sciences, Beijing Normal University, Zhuhai 519087, China}
\affiliation{$^e$Department of Physics and Synergistic Innovation Center for Quantum Effects and Applications, Hunan Normal University, Changsha, Hunan 410081, China}
\affiliation{$^f$Henan Academy of Sciences, Zhengzhou 450046, Henan, China}
\affiliation{$^g$College of Mathematics and Physics, Hunan University of Arts and Science, Changde, 415000, China}

\emailAdd{jinjiaheng@itp.ac.cn}
\emailAdd{zucheng.chen@bnu.edu.cn}
\emailAdd{yz@bnu.edu.cn}
\emailAdd{you\_zhiqiang@whu.edu.cn}
\emailAdd{liulang@bnu.edu.cn}
\emailAdd{youwuphy@gmail.com}

\abstract{
The stochastic signal detected by pulsar timing arrays (PTAs) has raised great interest in understanding its physical origin. 
Assuming the signal is a cosmological gravitational-wave background produced by overly large primordial curvature perturbations, we investigate the sound speed resonance effect with an oscillatory behavior using the combined PTA data from NANOGrav 15-yr data set, PPTA DR3, and EPTA DR2. 
We find that the stochastic signal can be explained by the induced gravitational waves sourced by the sound speed resonance mechanism, with the oscillation frequency $f_* \in [1.51, 4.90] \times 10^{-7}$Hz and the start time of oscillation $|\tau_0| \in [2.05, 106] \times 10^7$s.
}

\maketitle

\section{Introduction}

Recently, four regional pulsar timing array (PTA) collaborations, including the North American Nanohertz Observatory for Gravitational Waves (NANOGrav) \cite{NANOGrav:2023gor,NANOGrav:2023hde}, the Parkes PTA (PPTA) \cite{Reardon:2023gzh,Zic:2023gta}, the European PTA (EPTA) \cite{Antoniadis:2023rey,Antoniadis:2023utw}, and the Chinese PTA (CPTA) \cite{Nan:2011um,Xu:2023wog}, have announced the evidence of a stochastic signal consistent with the Hellings-Downs~\cite{Hellings:1983fr} spatial correlations in their latest data sets, pointing to the gravitational wave (GW) origin 
of this signal. Although many physical phenomena~\cite{Li:2019vlb,Chen:2021wdo,Wu:2021kmd,Chen:2021ncc,Chen:2022azo,PPTA:2022eul,Wu:2023pbt,Wu:2023dnp} can be the source in the PTA frequency band, the exact source of this signal, whether originating from supermassive black hole binaries~\cite{NANOGrav:2023hfp,Antoniadis:2023xlr,Bi:2023tib} or other cosmological sources~\cite{Vagnozzi:2023lwo,Han:2023olf,Li:2023yaj,Franciolini:2023wjm,Shen:2023pan,Kitajima:2023cek,Franciolini:2023pbf,Addazi:2023jvg,Cai:2023dls,Inomata:2023zup,Wang:2023ost,Murai:2023gkv,Li:2023bxy,Anchordoqui:2023tln,Liu:2023ymk,Abe:2023yrw,Ghosh:2023aum,Figueroa:2023zhu,Yi:2023mbm,Wu:2023hsa,Li:2023tdx,Geller:2023shn,You:2023rmn,Antusch:2023zjk,Ye:2023xyr,HosseiniMansoori:2023mqh,Liu:2023pau}, remains the subject of ongoing investigation. 

Scalar-induced gravitational waves (SIGWs)~\cite{Ananda:2006af,Baumann:2007zm,Garcia-Bellido:2016dkw,Inomata:2016rbd, Garcia-Bellido:2017aan,Kohri:2018awv,Cai:2018dig,Lu:2019sti,Yuan:2019wwo,Chen:2019xse,Yuan:2019fwv,Sakharov:2021dim}, accompanying the formation of primordial black holes (PBHs) \cite{Zeldovich:1967lct,Hawking:1971ei,Carr:1974nx,Saito:2008jc,Belotsky:2014kca,Carr:2016drx,Garcia-Bellido:2017mdw,Carr:2017jsz,Germani:2017bcs,Chen:2018czv,Chen:2018rzo,Liu:2018ess,Liu:2019rnx,Chen:2019irf,Yuan:2019udt,Fu:2019ttf,Cai:2019elf,Liu:2019lul,Cai:2019bmk,Liu:2020cds,Wu:2020drm,Fu:2020lob,Yi:2020kmq,DeLuca:2020sae,Vaskonen:2020lbd,DeLuca:2020agl,Domenech:2020ers,Yi:2020cut,Hutsi:2020sol,Yi:2021lxc,Kawai:2021edk,Chen:2021nxo,Braglia:2021wwa,Liu:2021jnw,Braglia:2022icu,Ashoorioon:2022raz,Chen:2022fda,Yi:2022anu,Yi:2022ymw,Chen:2022qvg,Liu:2022iuf,Zheng:2022wqo,Guo:2023hyp}, are a promising candidate for explaining this observed signal.
The SIGWs are produced by the large curvature perturbations generated during the inflationary epoch.   
Recently, a novel mechanism for PBH formation and the production of SIGWs, known as sound speed resonance (SSR), was proposed in \cite{Cai:2018tuh,Cai:2019jah,Chen:2019zza,Chen:2020uhe,Zhou:2020kkf}. The SSR mechanism involves the presence of an oscillating square of the sound speed, which gives rise to a nonperturbative parametric amplification of specific perturbation modes during the inflationary epoch. Consequently, the power spectrum of primordial curvature perturbations exhibits some pronounced peaks at small scales, while maintaining near-scale invariance on larger scales, consistent with predictions from inflationary cosmology. It is noteworthy that this novel mechanism predicts the existence of several narrow peaks in the power spectrum of primordial curvature perturbations at smaller scales. Additionally, in accordance with the second-order cosmological perturbation theory, the enhanced primordial curvature perturbations from SSR mechanism are expected to induce significant GW signals. 


In this work, assuming that the signal detected by PTAs originates from SIGWs, we jointly use the NANOGrav 15-yr data set, PPTA DR3, and EPTA DR2 to constrain the SSR model. In particular, we aim to explore whether the PTA signal is consistent with the SSR mechanism and also constrain the parameter space of SSR model. The rest of the paper is organized as follows. In Section \ref{SSR}, we provide an overview of the SSR model, focusing on the generation of primordial curvature perturbations. 
In Section \ref{IGW}, we introduce the energy density of SIGWs and PBH production.
In Section \ref{data}, we describe the methodology for data analyses and present the results obtained from the latest PTA data sets.
Finally, we make the conclusions in Section \ref{Con}. 

\section{Sound speed resonance}
\label{SSR}

In our paper, we will briefly review the mechanisms underlying SSR. To begin, we introduce the variable $u$, defined as $u \equiv z \zeta$, where $z \equiv \sqrt{2 \epsilon} a / c_s$. 
Here, $a$ denotes the scale factor, $c_s$ represents the sound speed, and $\epsilon$ is the slow-roll parameter defined as $\epsilon \equiv - \dot H / H^2$. 
Additionally, $\zeta$ represents the comoving curvature perturbation, which remains invariant in the gauge.
Considering the perturbation of Einstein equations, the Fourier mode of the variable $u$ satisfies the following equation \cite{Armendariz-Picon:1999hyi,Garriga:1999vw}
\begin{equation}\label{eq:uk}
    u''_{k} + \left( c_s^2 k^{2} - \frac{z''}{z} \right) u_{k} = 0,
\end{equation}
where the prime denotes a derivative with respect to the conformal time $\tau$. In the SSR model, the oscillating sound speed is defined as follows \cite{Cai:2018tuh}
\begin{equation}\label{eq:cs}
    c_s^2 = \left\{
    \begin{aligned}
        &1, &\tau < \tau_0,\\
        &1 -2 \xi \left[ 1 - \cos \left( 2 k_* \tau \right) \right], &\tau > \tau_0,
    \end{aligned}
    \right.
\end{equation}
where $\xi$ represents the amplitude of the oscillation, $k_*$ corresponds to the oscillation frequency, $\tau_0$ is the start time of oscillation. 
The oscillating sound speed can be realized within the effective field theory of inflation \cite{Achucarro:2010da, Achucarro:2012sm}. 
For instance, in the multi-field inflation model, the sound speed is given by $c_s^2 = (1+4\eta_\perp^2 H^2/M_0^2)^{-1}$.
Here, $M_0$ has the dimension of mass and  $\eta_\perp = V_N/(\dot{\phi_0}H)$, where $\phi_0$ and $V_N$ are related to the background inflatons and the potential, respectively~\cite{Achucarro:2010da}. 
Consequently, if there are oscillations in the potential of the inflation model, the oscillating sound speed in Eq. \eqref{eq:cs} can be obtained.

It is important to ensure that $c_s^2 \geq 0$, which imposes the condition $\xi < 1/4$. 
Additionally, $k_*\tau_0$ is chosen to be a multiple of $\pi$ to ensure a smooth transition of the sound speed from its constant value to the oscillating regime. 
As a result, the effective mass during the oscillation period can be simplified as
\begin{equation}
    \frac{z''}{z} = \frac{2}{\tau^2} - \frac{4 \xi k_*}{\tau}\sin(2 k_* \tau) + 4 \xi k_*^2 \cos(2 k_* \tau)+\mathcal{O}(\xi^2).
\end{equation}
Focusing on the sub-Hubble radius situation where $|k_* \tau| \gg 1$, the effective mass simplifies to $z''/z = 4 \xi k_*^2 \cos(2 k_*\tau)$.
The perturbation equation \eqref{eq:uk} can be rewritten as
\begin{equation}
    \frac{d^2 u_k}{dx^2} + \left( A_k - 2q \cos 2x \right) u_k =0,
\end{equation}
where we introduce the variable $x \equiv - k_* \tau$, $A_k = (1 - 2 \xi) k^2 / k_*^2$ and $q = (2 - k^2 / k_*^2) \xi$. 
This equation resembles the Mathieu equation, which exhibits parametric instability within specific ranges of $k$.
The resonance bands are located in narrow ranges around $k \simeq n k_*$, given that $\xi \ll 1$ and $|q| \ll 1$.

Let's consider the first resonance band with $n = 1$ as an example.
During the onset of resonance, we set the initial value of $u_k$ to the Bunch-Davies vacuum as $u_k(\tau_0)=e^{-i k \tau_0}/ \sqrt{2k}$.
For modes outside the resonance bands, $u_k$ remains approximately constant within the Hubble radius. 
However, for modes within the resonance bands, the behavior of $u_k$ inside the Hubble radius can be described by
\begin{equation}\label{eq:uk_tau}
    u_k(\tau) \propto \exp (\xi k_* \tau /2).
\end{equation}
Using $u \equiv z \zeta$ and Equation \eqref{eq:uk_tau}, the evolution of the comoving curvature perturbation for the $k_*$ mode can be approximated as
\begin{equation}
    \zeta_{k_*}(\tau) \simeq \zeta_{k_*}(\tau_0) e^{ \xi k_* (\tau - \tau_0)/2} \frac{\tau}{\tau_0}.
\end{equation}
This approximation holds within the Hubble radius, where $\zeta_{k_*}(\tau_0) = -H \tau_0 /\sqrt{4 \epsilon k_*}$, which is determined by the Bunch-Davies vacuum.
When the mode crosses the Hubble radius at $\tau_*=-1/k_*$, the amplitude is enhanced to
\begin{equation}
    \zeta_{k_*} \simeq \zeta_{k_*}(\tau_0) \left( \frac{-1}{k_* \tau_0} \right) e^{-\xi k_*\tau_0/2} \simeq \frac{H}{\sqrt{4 \epsilon k_*^3}} e^{-\xi k_* \tau_0/2}.
\end{equation}

The power spectrum of the primordial curvature perturbation, denoted as $P_\zeta$, is defined as $P_\zeta \equiv k^3 |\zeta_k|^2 /(2 \pi^2)$. Here, the resonance frequency exhibits an exponential amplification. As a result, the complete power spectrum can be parametrized as
\begin{equation}\label{Pk0}
    P_\zeta(k) = A_s \left( \frac{k}{k_p} \right)^{n_s-1} \left\{ 1+\frac{\xi k_*}{2}e^{-\xi k_*\tau_0} \left[ \delta(k-k_*) + \sum_{n=2}^{\infty} a_n \delta(k - n k_*) \right] \right\},
\end{equation}
where $A_s = H^2 / (8 \pi^2 \epsilon) \simeq 2.2\times 10^{-9}$ represents the amplitude of the power spectrum as in the standard inflation, $n_s \simeq 0.965$ corresponds to the spectral index at the pivot scale $k_p \simeq 0.05 ~ \text{Mpc}^{-1}$ \cite{Planck:2018vyg}, and $a_n \ll 1$ represents the amplitude of the $n$th peak relative to the first peak.
The coefficient in front of the $\delta$-function is determined by using a triangle approximation to estimate the area of the peak. Throughout our paper, we only consider the second and third peaks which are located at $2k_*$ and $3k_*$, respectively, and ignore other higher order peaks. This choice is motivated by the fact that the amplitudes of the peaks at higher orders are exponentially suppressed. Therefore we truncated the series at $n=3$ to ease the computational burden.

\section{SIGWs and PBHs}
\label{IGW}

In this section, we will introduce the formalism of SIGWs and PBHs. In the Newtonian gauge, the perturbed metric can be expressed as
\begin{equation}
d s^2=a^2(\tau) \left\{ -(1 + 2\Phi) d \tau^2 + \left[ (1 - 2\Phi) \delta_{ij} + \frac{1}{2} h_{ij} \right] d x^i d x^j \right\},
\end{equation}
where $\Phi$ represents the Bardeen potential, and $h_{ij}$ represents the tensor perturbations. Here, we neglect the effects of first-order gravitational waves, vector perturbations, and anisotropic stress, as previous studies (\cite{Baumann:2007zm,Weinberg:2003ur,Watanabe:2006qe}) have demonstrated that their contributions are small.
According to the Einstein field equations, the tensor perturbations in Fourier space can be written as \cite{Ananda:2006af,Baumann:2007zm}
\begin{equation}
h''_{\boldsymbol{k}}+2\mathcal{H} h'_{\boldsymbol{k}}+k^2 h_{\boldsymbol{k}} = 4S_{\boldsymbol{k}},
\end{equation}
where $\mathcal{H} \equiv  a'/a$  is the comoving Hubble parameter, and $S_{\boldsymbol{k}}$ represents the source term given by
\begin{equation}
S_{\boldsymbol{k}} = \int \frac{d^3 \boldsymbol{p}}{(2\pi)^{3/2}} e_{ij}(\boldsymbol{k}) p^i p^j \left[ 3\Phi_{\boldsymbol{p}}\Phi_{\boldsymbol{k}-\boldsymbol{p}} + \mathcal{H}^{-1} \left( \Phi’_{\boldsymbol{p}} \Phi_{\boldsymbol{k}-\boldsymbol{p}} + \Phi_{\boldsymbol{p}} \Phi’_{\boldsymbol{k}-\boldsymbol{p}} \right) + \mathcal{H}^{-2} \Phi’_{\boldsymbol{p}} \Phi’_{\boldsymbol{k}-\boldsymbol{p}} \right].
\end{equation}
The Bardeen potential in the Fourier space,  denoted as $\Phi_{\boldsymbol{k}}$ , is related to the primordial curvature perturbations $\zeta_{\boldsymbol{k}}$ as $\Phi_{\boldsymbol{k}}=2/3 ~\zeta_{\boldsymbol{k}}$. The solution for tensor perturbations is given by
\begin{equation}\label{hk:green}
h_{\boldsymbol{k}}(\tau)=\frac{4}{a(\tau)}\int^{\tau}d \tau’ g_{\boldsymbol{k}}(\tau,\tau’) a(\tau’) S_{\boldsymbol{k}}(\tau’),
\end{equation}
where the Green function is defined as
\begin{equation}
g_{\boldsymbol{k}}(\tau,\tau')=\frac{\sin \left[ k(\tau-\tau') \right]}{k}.
\end{equation}
The power spectrum of tensor perturbations is given by:
\begin{equation}
\langle h_{\boldsymbol{k}}(\tau) h_{\boldsymbol{k}'}(\tau)\rangle = \frac{2 \pi^2}{k^3} \delta^{3}(\boldsymbol{k}+\boldsymbol{k}') \mathcal{P}_h(k,\tau).
\end{equation}
Hence, the power spectrum of tensor perturbations can be expressed as \cite{Ananda:2006af,Baumann:2007zm,Kohri:2018awv,Espinosa:2018eve}
\begin{equation}
\mathcal{P}_h(k,\tau) = 4 \int_{0}^{\infty}dv \int_{|1-v|}^{1+v} du \left[ \frac{4v^2 - \left( 1-u^2+v^2 \right)^2}{4uv}\right]^2 I_{RD}^2(u,v) \mathcal{P}_{\zeta}(kv) \mathcal{P}_{\zeta}(ku),
\end{equation}
where $u=|\boldsymbol{k}-\boldsymbol{p}|/k$, $v=p/k$, and $I_{RD}$ is the integral kernel. In the study of SGWBs, their characterization often involves quantifying the energy density per logarithmic frequency interval relative to the critical density $\rho_c(\tau)$, which is  expressed as:
\begin{equation}
\Omega_{\mathrm{GW}}(k, \tau) \equiv \frac{1}{\rho_c(\tau)} \frac{\mathrm{d} \rho_{\mathrm{GW}}(k, \tau)}{\mathrm{d} \ln k}=\frac{k^3}{48 \pi^2}\left(\frac{k}{\mathcal{H}}\right)^2 \overline{\left\langle\left|h_{\bm{k}}(\tau)\right|^2\right\rangle},
\end{equation}
Here, the term $\overline{\left\langle\left|h_{\bm{k}}(\tau)\right|^2\right\rangle}$ represents an average over a few wavelengths. During the radiation-dominated era, gravitational waves are generated by curvature perturbations. The density parameter of gravitational waves at the matter-radiation equality epoch is denoted as $\Omega_{\mathrm{GW}}(k)=\Omega_{\mathrm{GW}}(k,\tau\rightarrow\infty)$. Therefore, the energy density of GWs can be expressed as \cite{Espinosa:2018eve}
\begin{equation}
\Omega_{\mathrm{GW}}(k) = \int_0^{\infty} \mathrm{d} v \int_{|1-v|}^{|1+v|} \mathrm{d} u \mathcal{T}(u, v) {P}_{\zeta}(v k) {P}_{\zeta}(u k),
\end{equation}
with 
\begin{equation}
\begin{aligned}
&\mathcal{T}(u,v)=  \frac{3}{1024 v^8 u^8}\left[4 v^2-\left(v^2-u^2+1\right)^2\right]^2\left(v^2+u^2-3\right)^2 \\
& \times\left\{\left[\left(v^2+u^2-3\right) \ln \left(\left|\frac{3-(v+u)^2}{3-(v-u)^2}\right|\right)-4 v u\right]^2+\pi^2\left(v^2+u^2-3\right)^2 \Theta(v+u-\sqrt{3})\right\}.
\end{aligned}
\end{equation}
Here, $\Theta$ is the Heaviside theta function. By employing the relationship between the wave number $k$ and frequency $f$, $k = 2\pi f$, we can derive the energy density fraction spectrum of SIGWs at the present time,
\begin{equation}
\Omega_{\mathrm{GW}, 0}(f)=\Omega_{\mathrm{r}, 0}\left[\frac{g_{*, r}(T)}{g_{*, r}\left(T_{\mathrm{eq}}\right)}\right]\left[\frac{g_{*, s}\left(T_{\mathrm{eq}}\right)}{g_{*, s}(T)}\right]^{\frac{4}{3}} \Omega_{\mathrm{GW}}(k),
\end{equation}
where $g_{*,s}$ and $g_{*,r}$ are the effective degrees of freedom for entropy density and radiation, and $\Omega_{r,0}$ denotes the present energy density fraction of radiation.

{PBHs are produced from gravitational collapse when the density contrast $\delta\rho/\rho$ exceeds a critical threshold $\delta_c$ within Hubble patches. The exact value of $\delta_c$ depends on the equation of state parameter $w$ and the propagation speed $c_s^2$ ~\cite{Zeldovich:1967lct,Hawking:1971ei,Carr:1974nx,Meszaros:1974tb,Carr:1975qj,Musco:2004ak,Musco:2008hv,Musco:2012au,Harada:2013epa,Escriva:2020tak}.  In a cautious estimation, we adopt Carr's criterion in the uniform Hubble slice~\cite{Carr:1975qj}, which can be translated to the comoving slice as~\cite{Domenech:2020ers}
\begin{align}\label{dcb}
\delta_c \simeq \frac{3(1+w)}{5+3w} c_s^2.
\end{align}
The rationale for our selection of $\delta_c$ is justified by the fact that the fluctuations of the scalar field propagate at the speed of sound squared, making PBH formation more challenging than in an adiabatic perfect fluid. Thus, we adopt the upper limit of the density threshold from Ref.~\cite{Harada:2013epa}. When $\xi \ll 1$, $c_s^2$ oscillates from $1-4\xi$ to $1$, which is  nearly a constant. Therefore, in SSR model and the radiation-dominated era, $\delta_c \approx 2/3$, which can alleviate the tension of PBH overproduction. The PBH abundance is highly sensitive to the power spectrum, with the first band (n = 1) experiencing much greater enhancement than subsequent harmonic bands. Consequently, in the PBH abundance calculation, our focus is solely on the resonated modes around $k_*$. The relation of PBH mass $M$ and wavenumber $k$ is given by
\begin{equation}
k \approx \frac{1.7 \times 10^7}{\mathrm{Mpc}} \left(\frac{M_{\odot}}{M}\right)^{1 / 2}\left(\frac{\gamma}{0.2}\right)^{1/2} \left(\frac{g_{*, r}\left(T\left(M\right)\right)}{106.75}\right)^{1 / 4}\left(\frac{g_{*, s}\left(T\left(M\right)\right)}{106.75}\right)^{-1 / 3},
\end{equation}
where $M_{\odot}$ is the solar mass and $\gamma\approx0.2$ denotes the fraction of matter within the Hubble horizon that undergoes gravitational collapse, leading to the formation of PBHs.
}

{The estimation of PBH abundance with mass $M(k_*)$ typically involves defining $\beta(M)$ as the mass fraction of PBHs relative to the total energy density at the time of formation. This quantity can be expressed as an integration of the Gaussian distribution of perturbations
\begin{equation}
    \beta(M)=\frac{\gamma}{2}\text{erfc}\left(\frac{\delta_c}{\sqrt2\sigma(M)}\right).
\end{equation}
The quantity $\sigma(M)$ denoting the variance of the density perturbation smoothed over the mass scale of $M$, is estimated as
\begin{equation}
\label{sigma}
\sigma^{2} =\frac{16}{81} ~ \int_{0}^{\infty}\! \frac{\mathrm{d}q}{q}\, \left(\frac{q}{k}\right)^{4} \tilde{W}^{2}(q/k) T^{2}(q/k) \mathcal{P}_{\mathcal{R}}(q),
\end{equation}
where the function $\tilde{W}(q/k)=\exp(-q^{2}/2k^{2})$ corresponds to the Gaussian window function, and $T(q/k)=3(\sin l - l \cos l)/l^{3}$ is the transfer function, with $l = q/(\sqrt{3}k)$. When the primordial scalar power spectrum exhibits a very sharp peak, it leads to a monochromatic mass function for the PBHs. The  total abundance of PBHs in the dark matter at present can be defined as \cite{Sasaki:2018dmp}
\begin{equation}
f_{\mathrm{PBH}}\equiv \frac{\Omega_{\mathrm{PBH}}}{\Omega_{\mathrm{CDM}}} = 2.7 \times 10^8 \left(\frac{g_{*,r}}{10.75}\right)^{3/4} \left(\frac{g_{*,s}}{10.75}\right)^{-1} \left(\frac{M}{M_{\odot}}\right)^{-1/2} \beta(M),
\end{equation}
where $\Omega_{\mathrm{CDM}}$ represents the density of cold dark matter. 
}

\begin{table}
    \centering
	\begin{tabular}{c|ccc}
		\hline\hline
		Parameter & $\log_{10} (f_*/\mathrm{Hz})$ & $\log_{10} \xi$ & $\log_{10} (-\tau_0/\mathrm{s})$\\
		\hline
		 Prior& \quad $\mU(-9, -5)$\quad & \quad $\mU(-4, \log_{10} 0.25)$\quad & \quad $\mU(5, 12)$\quad\\[1pt]
		NANOGrav result & $-5.85^{+0.77}_{-1.17}$ & $-2.17^{+1.43}_{-1.62}$ & \quad $8.66^{+2.07}_{-1.81}$\\[1pt]
		PPTA result & $-5.83^{+0.74}_{-0.80}$ & $-2.13^{+1.38}_{-1.64}$ & \quad $8.55^{+1.92}_{-1.70}$\\[1pt]
		EPTA result & $-6.86^{+1.57}_{-0.76}$ & $-2.18^{+1.45}_{-1.60}$ & \quad $9.42^{+2.04}_{-2.00}$\\[1pt]
		Combined result & $-6.31^{+0.92}_{-0.51}$ & $-2.20^{+1.43}_{-1.57}$ & \quad $9.03^{+1.82}_{-1.71}$\\[1pt]
  \hline
	\end{tabular}
	\caption{\label{tab:priors}Prior distributions and results for the model parameters. We present the results obtained from the NANOGrav 15-yr data set, the PPTA DR3, the EPTA DR2, and the combined data (NANOGrav+PPTA+EPTA), respectively. Here $\mU$ denotes the uniform distribution. We quote each parameter's median value and $90\%$ equal-tail credible interval.}
\end{table}

\begin{figure}[htbp!]
	\centering
    \includegraphics[width=\textwidth]{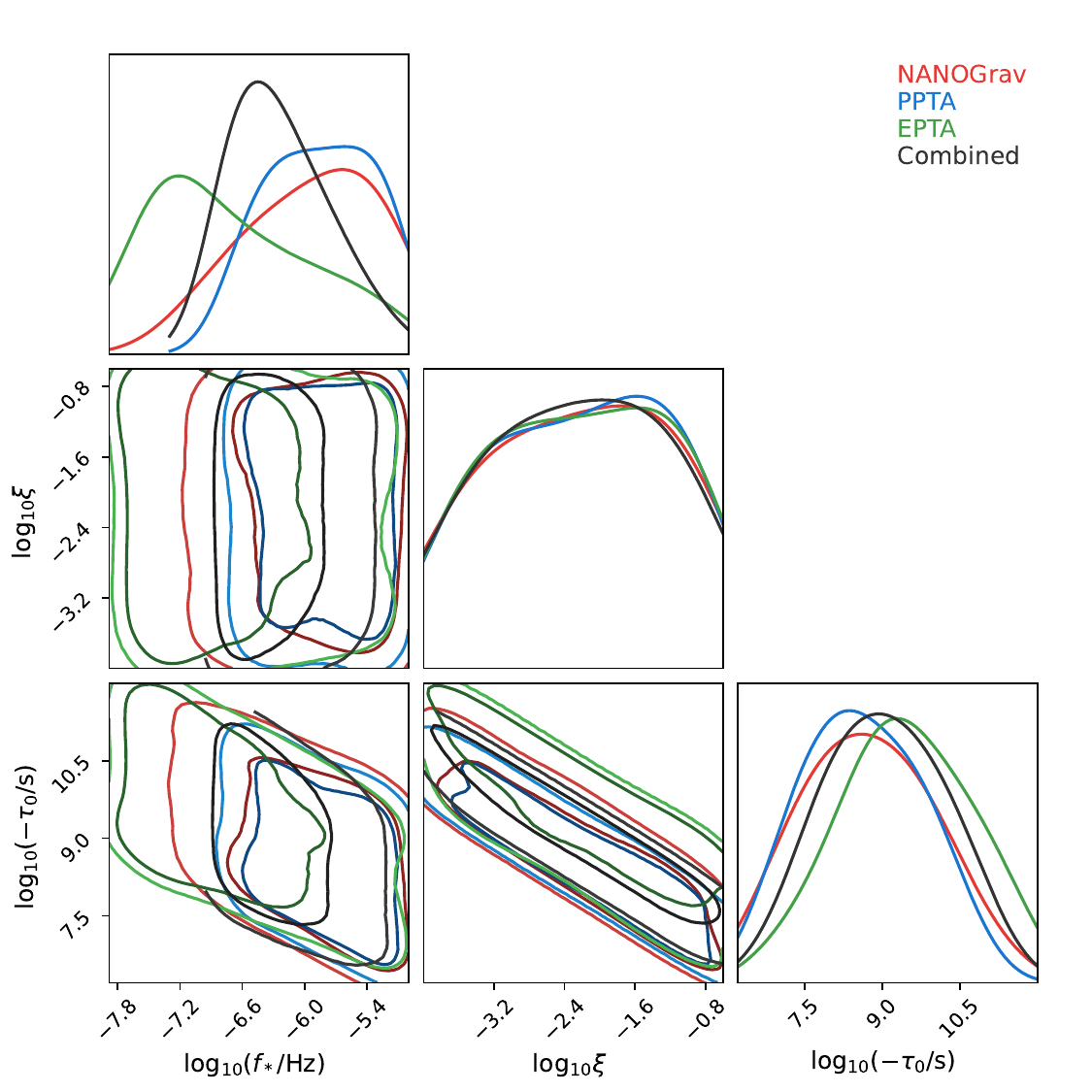}
	\caption{\label{posts_SSR}One and two-dimensional marginalized posteriors of the parameters obtained from the NANOGrav 15-yr data set, PPTA DR3, EPTA DR2, and the combined data (NANOGrav+PPTA+EPTA). The contours in the two-dimensional plot correspond to the $1 \sigma$ and $2 \sigma$ credible regions, respectively.}
\end{figure}

\begin{figure}[htbp!]
	\centering
	\includegraphics[width=\textwidth]{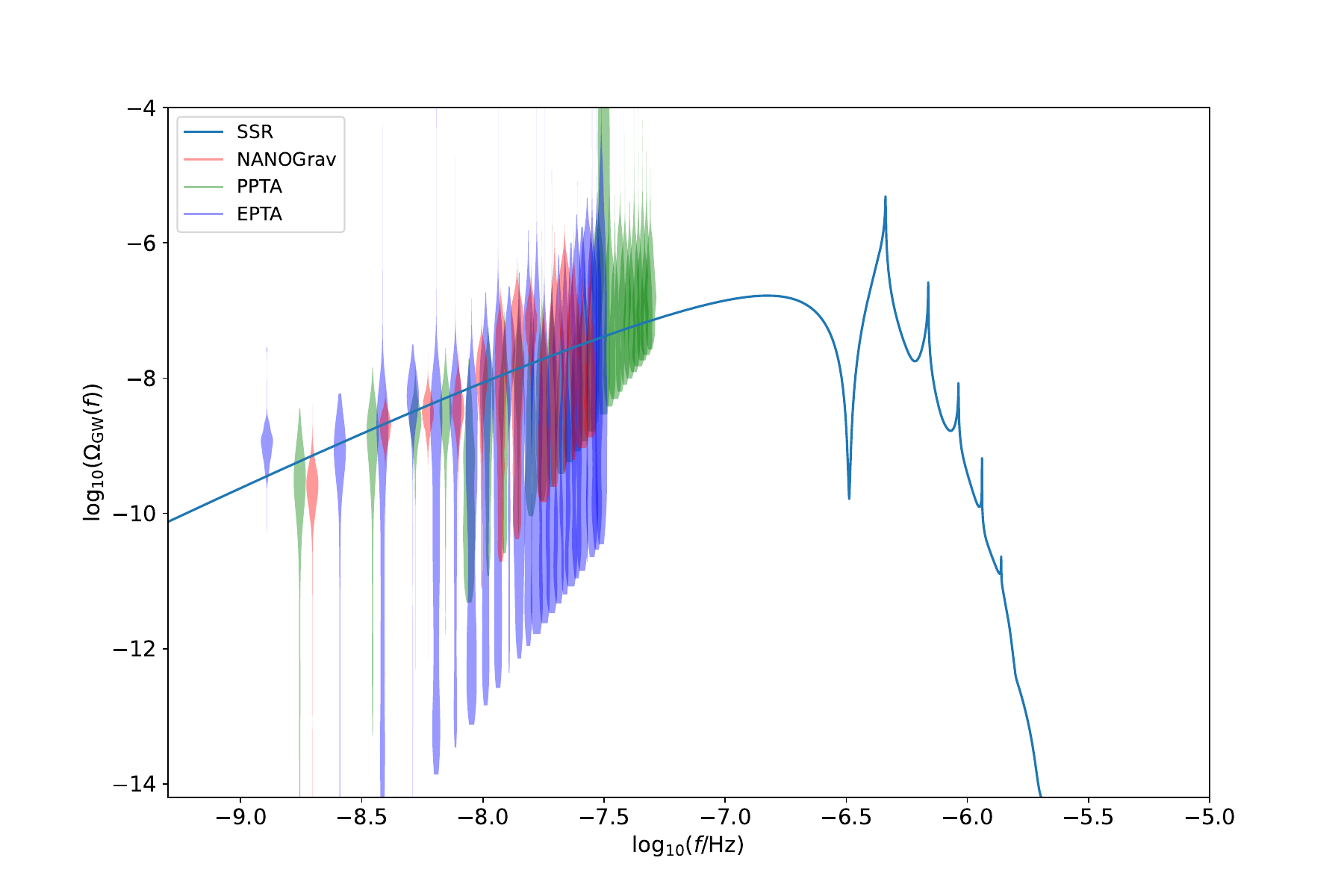}
	\caption{\label{ogw}The energy density of SIGWs sourced by the SSR effect. We also show the energy density spectra derived from the free spectrum from NANOGrav 15-yr data set (red violins), PPTA DR3 (green violins), and EPTA DR2 (blue violins). }
\end{figure}

\begin{figure}[htbp!]
	\centering
	\includegraphics[width=\textwidth]{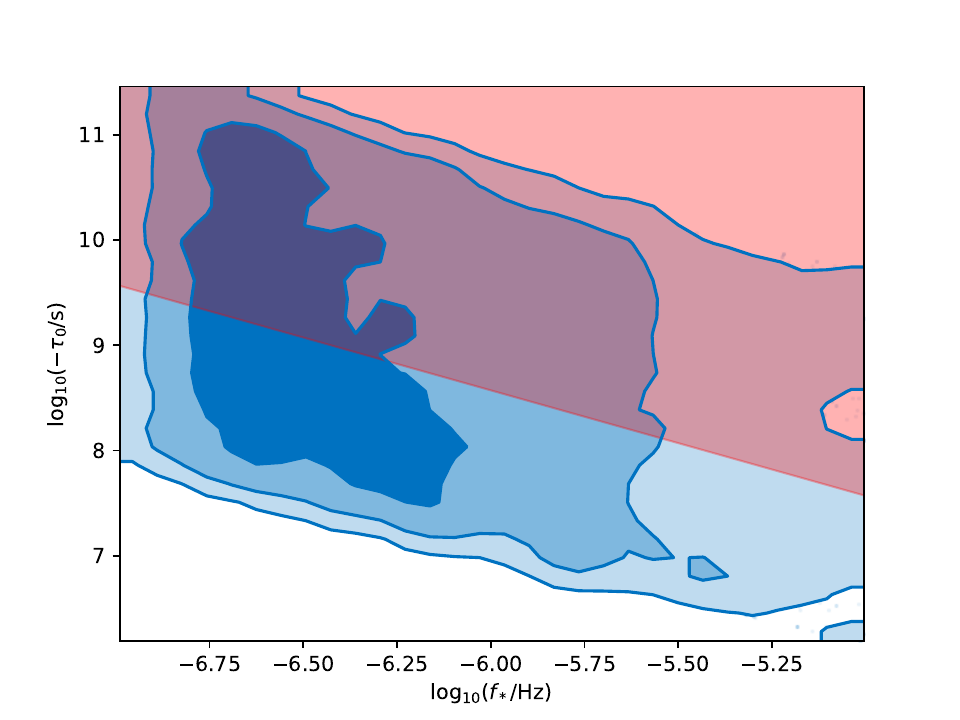}
	\caption{\label{post_with_fpbh}{The two dimensional posteriors for the $\log_{10} (f_*/\mathrm{Hz})$ and $\log_{10} (-\tau_0/\mathrm{s})$ parameters shown in blue contours obtained from the combined data (NANOGrav+PPTA+EPTA). The red region is the excluded parameter space where $\fpbh > 1$ by fixing $\xi=0.01$.}}
\end{figure}

\section{Data analysis and results}
\label{data}
Following \cite{Liu:2023ymk,Wu:2023hsa}, we jointly use the NANOGrav 15-yr data set, the PPTA DR3, and the EPTA DR2 to estimate the model parameters. In this work, we use the free spectrum data obtained from each PTA with Hellings-Downs~\cite{Hellings:1983fr} spatial correlations. Given the time span $T_{\mathrm{obs}}$ of a PTA, the free-spectrum starts with the lowest frequency $1/T_{\mathrm{obs}}$. NANOGrav, PPTA, and EPTA use $14$~\cite{NANOGrav:2023gor}, $28$~\cite{Reardon:2023gzh}, and $24$~\cite{Antoniadis:2023rey} frequencies in searching for the GWB signal, respectively. {In this work, we constrain the parameter space by considering the NANOGrav 15-yr data set, the PPTA DR3, the EPTA DR2, and the combined data (NANOGrav+PPTA+EPTA), respectively. In particular, the combined data consists $66$ frequencies for a free spectrum ranging from $1.28$~nHz to $49.1$~nHz. We note that although the simply combined data can improve the sensitivity in searching for SGWB signal, a proper way to combine the different PTA data sets should wait for the IPTA DR3 which is undergoing.}

We use the free-spectrum time delay data released by each PTA. The time delay $d(f)$ is related to the power spectrum $S(f)$ by
\begin{equation}
d(f)=\sqrt{S(f) / T_{\mathrm{obs}}}.
\end{equation}
We then obtain the characteristic strain, $h_c(f)$, by
\begin{equation}
h_c^2(f)=12 \pi^2 f^3 S(f).
\end{equation}
Finally, we can calculate the free-spectrum energy density as
\begin{equation}
\hat{\Omega}_{\mathrm{GW}}(f)=\frac{2 \pi^2}{3 H_0^2} f^2 h_c^2(f) = \frac{8\pi^4}{H_0^2} T_{\mathrm{obs}} f^5 d^2(f).
\end{equation}
For each observed frequency, $f_i$, we can estimate the corresponding kernel density $\mathcal{L}_i$ given the posteriors of $\hat{\Omega}_{\mathrm{GW}}(f_i)$ derived above. Therefore, the total likelihood is~\cite{Liu:2023ymk,Wu:2023hsa}
\begin{equation}
    \mathcal{L}(\Lambda) = \prod_{i=1}  \mathcal{L}_i(\Omega_{\mathrm{GW}}(f_i, \Lambda)),
\end{equation}
where $\Lambda\equiv \{f_*, \xi, \tau_0\}$ is the collection of the model parameters and $f_*=k_*/2\pi$. Table~\ref{tab:priors} summarizes the model parameters and their priors. Note that we sample in the log space for all these three parameters. In practice, we use \texttt{dynesty} \cite{Speagle:2019ivv} sampler wrapped in the \texttt{Bilby} \cite{Ashton:2018jfp,Romero-Shaw:2020owr} package to explore the parameter space.

The posterior distributions of the model parameters, $\Lambda = \{f_*, \xi, \tau_0\}$, are shown in Figure~\ref{posts_SSR}. {The results for each model parameter with median and $90\%$ equal-tail credible interval obtained from different PTA data sets are summarized in Table~\ref{tab:priors}. We see that the results derived from NANOGrav, PPTA, and EPTA are consistent with each other, and the combined data indeed have a better constraint on the parameters especially for the $f_*$ parameter. In what follows, we should focus on the results from the combined data.}
In particular, to explain the PTA signal, we obtain constraints on the oscillation frequency $\log_{10} (f_*/\mathrm{Hz}) = -6.31^{+0.92}_{-0.51}$, the amplitude of the oscillation $\log_{10} \xi =-2.20^{+1.43}_{-1.57}$, and the start time of oscillation $\log_{10} (-\tau_0/s) = 9.03^{+1.82}_{-1.71}$ with $90\%$ equal-tail uncertainties. {Note that the SSR effect can alleviate the tension of PBH overproduction as shown in Figure~\ref{post_with_fpbh}. Because $c_s^2$ oscillates in the range of $[1-4\xi, 1]$, it can be well approximated by $c_s^2 \simeq 1$ when $\xi \ll 1$. Therefore, in the SSR model, we have $\delta_c \approx 2/3$ which is much larger than the value $\delta_c \approx2/9$ as in an adiabatic perfect fluid where $c_s^2=w=1/3$ (see Eq.~\eqref{dcb}), making PBH formation in the SSR model with $c_s^2 \simeq 1$ more challenging than in an adiabatic perfect fluid where $c_s^2=w=1/3$. This effect has already been carefully discussed in Ref~\cite{Balaji:2023ehk}.}

A representative of the SIGWs sourced by the SSR effect is shown in Figure~\ref{ogw}, where five spikes appear in the energy density spectrum with $n=3$. A general feature of the SSR is that the number of spikes in the energy density spectrum is equal to $2n-1$ where $n$ is the number of peaks of the power spectrum defined in Eq.~\eqref{Pk0}. Furthermore, the spikes of the energy density spectrum locate at $2 f_*/\sqrt{3}$, $3 f_*/\sqrt{3}$, $\cdots$, $2n f_*/\sqrt{3}$.

\section{Conclusions}\label{Con}

The stochastic signal detected by NANOGrav, PPTA, EPTA, and CPTA shows strong evidence of the Hellings-Downs spatial correlations, suggesting a GW origin of this signal. Assuming the signal is from the SIGWs induced by the primordial curvature perturbations, we constrain the SSR effect with combined PTA data from NANOGrav 15-yr data set, PPTA DR3, and EPTA DR2.

We find that the stochastic signal can be explained by the induced gravitational waves sourced by the SSR mechanism, with the oscillation frequency $f_* \in [1.52, 4.67] \times 10^{-7}$Hz and the start time of oscillation $|\tau_0| \in [2.17, 95.7] \times 10^7$s. The relatively high oscillation frequency $f_*$ implies that the oscillation structure from SSR effect lies out of the PTA frequency band and cannot be probed by PTA.

The resonating primordial curvature perturbations with an oscillatory feature in the sound speed of their propagation may be realized in the context of an effective field theory of inflation or by non-canonical models inspired by string theory \cite{Cai:2018tuh}. We, therefore, expect the PTA data can also constrain the inflation models through the SSR mechanism. We leave it as future work.

\section*{Acknowledgments}
We thank the anonymous referee for providing constructive comments and suggestions that greatly improve the quality of this manuscript. ZCC is supported by the National Natural Science Foundation of China (Grant No. 12247176 and No. 12247112) and the China Postdoctoral Science Foundation Fellowship No. 2022M710429. 
ZY is supported by the National Natural Science Foundation of China under Grant No. 12205015 and the supporting fund for young researcher of Beijing Normal University under Grant No. 28719/310432102.
ZQY is supported by the China Postdoctoral Science Foundation Fellowship No. 2022M720482.
LL is supported by the National Natural Science Foundation of China (Grant No. 12247112 and No. 12247176) and the China Postdoctoral Science Foundation Fellowship No. 2023M730300.

\bibliographystyle{JHEP}
\bibliography{ref}

\end{document}